\documentclass[12pt]{article}
\usepackage{authblk}
\usepackage{hyperref}
\usepackage[letterpaper,margin=1in]{geometry}
\usepackage{graphicx}
\usepackage{lipsum, wrapfig}
\usepackage[labelfont=bf]{caption}
\usepackage{sectsty}
\usepackage{indentfirst}
\usepackage{titlesec}
\usepackage{amsmath}
\usepackage[normalem]{ulem}
\usepackage{multicol}
\usepackage{caption}
\begin{document}

\title{How the other half lives: \\
CRISPR-Cas's influence on bacteriophages}

\author
       {Melia E.\ Bonomo$^{1,3}$ and Michael W.\ Deem$^{1,2,3}$
       \\
       {\small $^1$\emph{Department of Physics and Astronomy, Rice University, Houston, TX 77005, USA}}\\
      {\small $^2$\emph{Department of Bioengineering, Rice University, Houston, TX 77005, USA}}\\
       {\small $^3$\emph{Center for Theoretical Biological Physics, Rice University, Houston, TX 77005, USA}}}

\date{}
\maketitle

\begin{abstract}
CRISPR-Cas is a genetic adaptive immune system unique to prokaryotic cells used to combat phage and plasmid threats.  The host cell adapts by incorporating DNA sequences from invading phages or plasmids into its CRISPR locus as spacers.  These spacers are expressed as mobile surveillance RNAs that direct CRISPR-associated (Cas) proteins to protect against subsequent attack by the same phages or plasmids.  The threat from mobile genetic elements inevitably shapes the CRISPR loci of archaea and bacteria, and simultaneously the CRISPR-Cas immune system drives evolution of these invaders.  Here we highlight our recent work, as well as that of others, that seeks to understand phage mechanisms of CRISPR-Cas evasion and conditions for population coexistence of phages with CRISPR-protected prokaryotes.
\end{abstract}

\section{Introduction}

Uncovering the structure, function, and potential applications of the prokaryotic CRISPR-Cas locus has been a growing research interest over the past 30 years~\cite{2}.  These loci contain a special family of {\bf c}lustered {\bf r}egularly {\bf i}nterspaced {\bf s}hort {\bf p}alindromic {\bf r}epeats (CRISPR) and a unique group of {\bf C}RISPR {\bf as}sociated (Cas) proteins encoded by \emph{cas} genes.  The 30-bp intervening sequences called `spacers' are of extrachromosomal origin and correspond to bacteriophage and plasmid genes, many of which are essential to infection or plasmid transference~\cite{40,41}.  Early discoveries from genomic sequence analyses, including the negative correlation found between the number of CRISPR spacers in \emph{Streptococcus thermophilus} and the strain's sensitivity to phage infection~\cite{122} and the lack of CRISPR loci in unthreatened laboratory strains, led researchers to postulate that these elements constituted a genetic adaptive im
 mune system 
 shaped by the host's immediate environment~\cite{180}.  Soon after, CRISPR-mediated phage resistance by the integration of spacers, as well as the loss of resistance following the deletion of these crucial spacers, was experimentally demonstrated~\cite{27,43}.

\begin{figure}
\begin{center}
\includegraphics[width=0.8\textwidth]{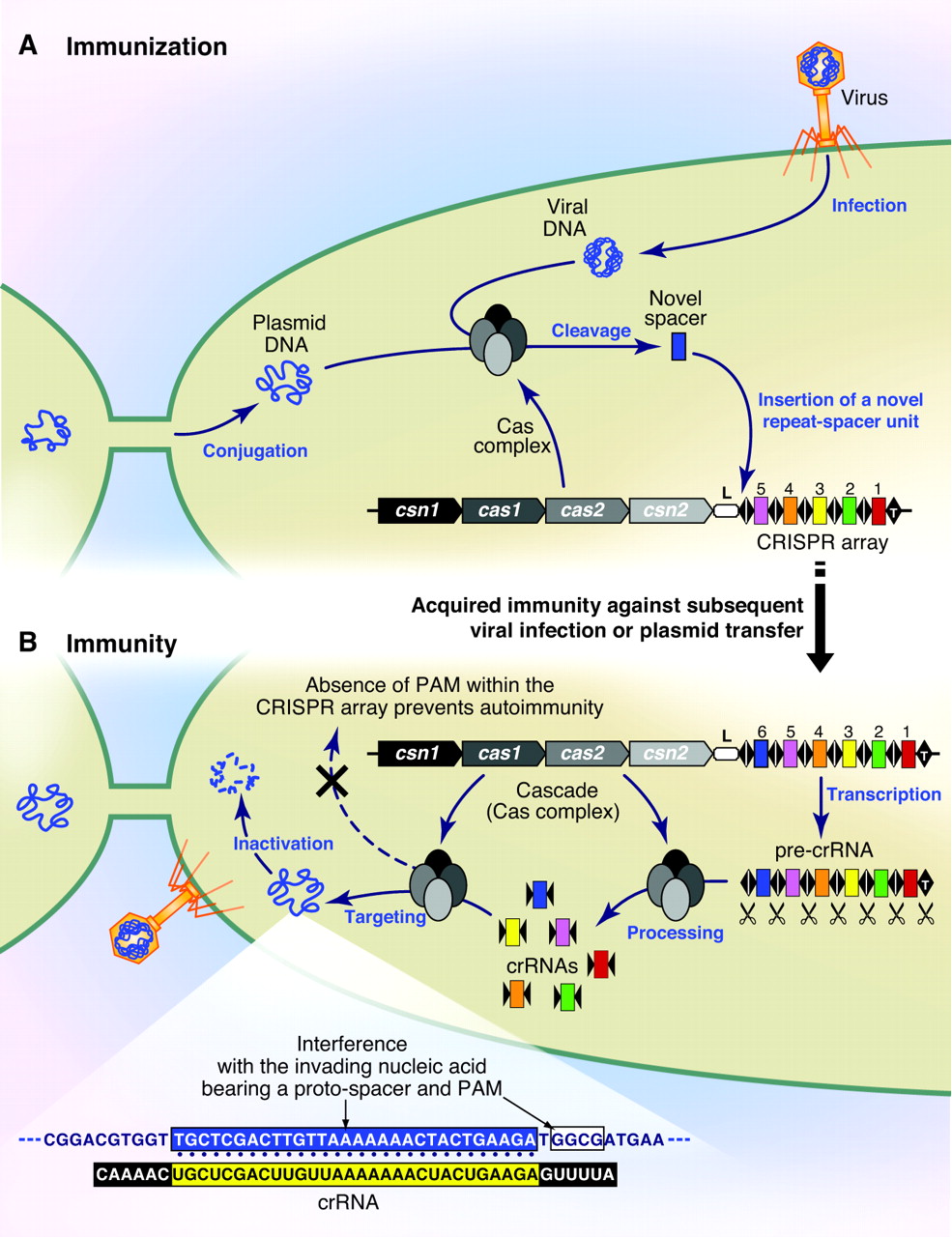}
\caption{The prokaryotic CRISPR-Cas defense cycles through three stages of adaptation, expression, and interference that are mediated by Cas proteins unique to the CRISPR locus (see also Figure~\ref{fig:casproteins}).  (A) During adaptation, the CRISPR-Cas system incorporates protospacer sequences from previous invaders into its locus as spacers.  A new repeat is copied as the spacer is inserted directly downstream from the leader sequence.  (B) During expression, a crRNA guide is created.  Depending on the type of CRISPR system, the crRNA is anchored either to one Cas protein or to a multi-component Cas protein complex.  During interference, the Cas protein(s) surveil mobile genetic elements that enter the cell and specifically cut sequences that match the crRNA to inhibit infection and replication.  There is experimental evidence of both DNA and RNA targeting, depending on the type of CRISPR-Cas system. Reprinted with permission from~\cite{76}.}
\label{fig:3stages}
\end{center}
\end{figure}

Though there is a vast variety of these systems~\cite{111}, the general mechanisms of CRISPR-Cas can be divided into three stages of adaptation, expression, and interference, as seen in Figure~\ref{fig:3stages}.  Biochemical and structural analyses have investigated the molecular mechanisms and conformational changes of the Cas proteins associated with each of these stages~\cite{55}.  The host cell combats phage and plasmid threats in its environment by encoding spacers into its genome from one or multiple DNA sequences of previous invaders, called protospacers.  New spacers are incorporated directly downstream of an AT-rich `leader' sequence, which characteristically flanks the start of the locus, and older spacers may be deleted at random.  The sequential ordering of spacer acquisition provides chronological information about the order in which a cell encountered each phage or plasmid.  Each spacer is then expressed as a mobile surveillance {\bf CR}ISPR {\bf RNA} (guide crR
 NA) that contains a single spacer and a partial repeat sequence on one or both sides.  In some CRISRP types, an additional {\bf tra}ns-activating {\bf CR}ISPR {\bf RNA} (tracrRNA) is needed to anchor the crRNA to the Cas surveillance protein.  The guide crRNA directs these Cas proteins to interfere with subsequent threats by targeting and specifically cleaving the invading DNA sequences, or in less common cases RNA sequences, that match those of the spacers.  Specificity requirements for the recognition of targets vary among CRISPRs.  Some require a perfect match between the guide crRNA and the target DNA sequences, while others can tolerate a certain number of mismatches if, for example, the invader has undergone a point mutation.  CRISPR-Cas systems that utilize a {\bf p}rotospacer {\bf a}djacent {\bf m}otif (PAM) to distinguish between self and target genomes generally cannot tolerate mutations in this motif region~\cite{25}.

\begin{figure}
\begin{center}
\includegraphics[width=0.9\textwidth]{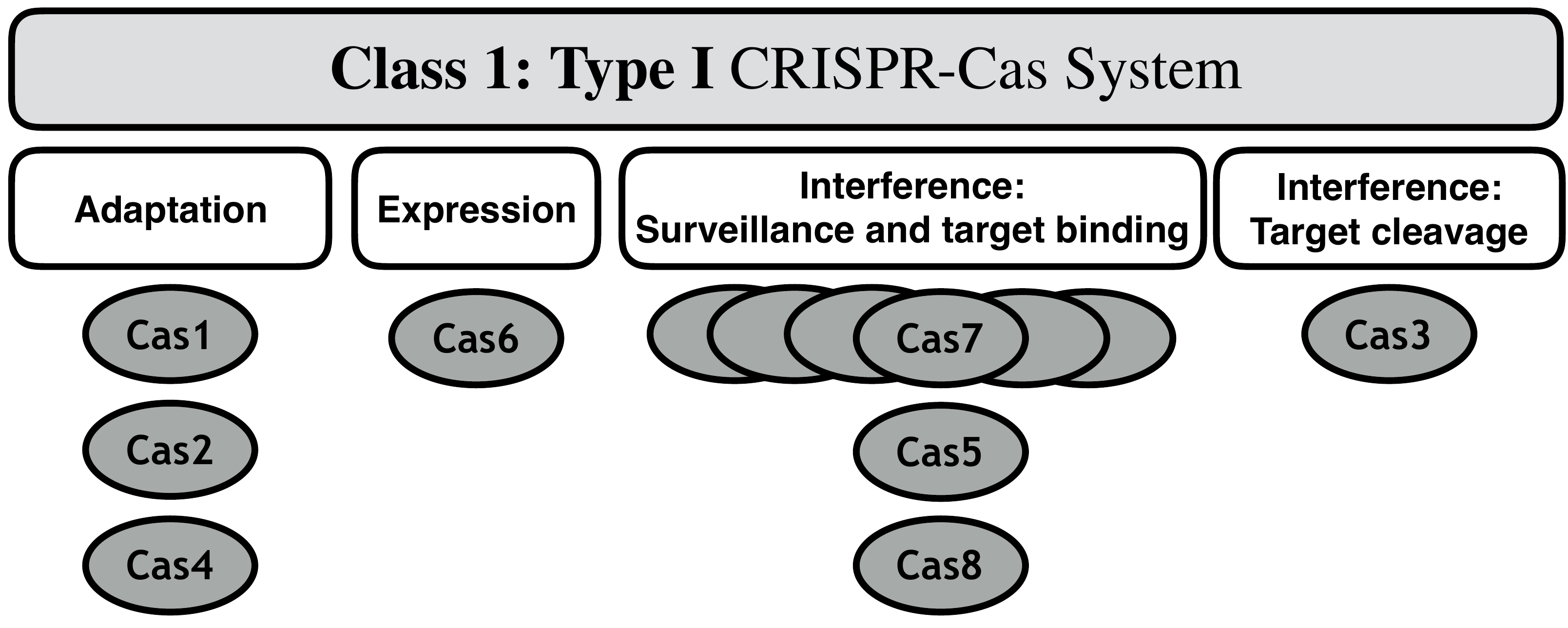}
 \caption{Each CRISPR-Cas system has a diverse set of Cas protein machinery.  For example, in the Class 1, Type I systems, Cas1, Cas2, and Cas4 are used to acquire spacers; Cas6 processes these spacers into crRNA; a complex of multiple Cas protein subunits is used to surveil the target sequence; and Cas3 is recruited for target cleavage.  The seven subtypes for Type I are I-A, I-B, I-C, I-D, I-E, I-F, and I-U. See Makarova \emph{et al.}, 2015, for the complete Cas protein content of other CRISPR systems~\cite{111}. }
\label{fig:casproteins}
\end{center}
\end{figure}

Most of the archaea and about half of the bacteria that have been sequenced contain functioning CRISPR-Cas systems.  An evolving CRISPR-Cas classification system~\cite{111} currently organizes these systems into two overarching classes, for those that utilize a single interference protein versus those that use multiple Cas protein units to survey and cut the target.  The systems are further delineated into six major types based on their principle \emph{cas} gene and more than 16 subtypes defined by their Cas protein content.  A single organism could have multiple types of CRISPR loci.  Additionally, there are still a number of rare, unclassified systems.  Figure~\ref{fig:casproteins} shows a representative example of the Cas protein content within the Type I CRISPR.

An extensive genomic analysis of the CRISPR repeats, spacers, leader sequences, and \emph{cas} genes in lactic acid bacteria genomes revealed the likelihood of CRISPR locus acquisition through horizontal gene transfer (HGT) between distant organisms~\cite{78}.  Interestingly, further HGT in a CRISPR-Cas-protected genome appears to be blocked, explaining the lack of loci in antibiotic resistant and lysogenic bacteria~\cite{81}.  Due to the polarized spacer acquisition that causes the ancestral end to contain phylogenetic anchors and the active end to contain recent encounters, the locus can be used to reconstruct the history of strain divergence~\cite{4}.  Additionally, CRISPR immunity has been shown to facilitate speciation within the \emph{Streptococcus}, \emph{Staphylococcus}, \emph{Lactobacillus}, and \emph{Bifidobacterium} genera~\cite{4}.  The loss of CRISPR-Cas in some strains within a given species allows those strains to acquire virulence via HGT, eventually leading t
 o the emergence of a new pathogenic species.  

In exploring how CRISPR-Cas systems could be manipulated, a locus was transferred from one organism into a distantly related one to confer protection against specific plasmids and phage infections~\cite{42}.  One of the more drastic transfers was an oral bacterium's RNA-targeting Type VI-A system that was successfully introduced into \emph{Escherichia coli}, which naturally contains DNA-targeting CRISPRs, to defend the cell from an RNA bacteriophage~\cite{169}.  Following this initial success, immunization of dairy industry-relevant prokaryotes was carried out to establish resistance to anticipated phage attacks~\cite{128,52}.  A turning point came in the applications side of the field when researchers realized the possibility of harnessing the CRISPR-Cas system to make specific genomic modifications in both prokaryotic and eukaryotic cells~\cite{55,D}.  The Cas9 protein from Type II systems can be re-programed with a single, custom guide sequence to make specific genomic cut
 s for sequence insertions or deletions.  Catalytically {\bf d}eactivated Cas9 (dCas9) can furthermore be fused to a promoter or represser to respectively {\bf a}ctivate (CRISPRa) or {\bf i}nterfere (CRISPRi) with targeted genes~\cite{1} and facilitate epigenetic studies~\cite{6}.

The coevolution of CRISPR-Cas-containing bacteria with plasmids and virulent phages creates what many call a coevolutionary ``arms race.''  Mathematical models that have been developed to investigate this coevolution either take a mean field approach to model the rate of change of population abundances, usually of wild-type and mutant phages and sensitive and immune bacteria, or look on a more detailed level at phage and bacterial strains represented as arrays of protospacers and spacers.  Generally, the degree of immunity, heritability, and benefit of maintaining the CRISPR-Cas system are studied as functions of the number and content of spacers, the abundance and diversity of phages and hosts, and CRISPR-associated fitness costs, such as autoimmunity and the restriction of HGT.  Koonin and Wolf, 2015, provides a detailed review of phage-host evolution models~\cite{7}.  While the threat from mobile genetic elements inevitably shapes the CRISPR loci of archaea and bacteria, i
 t is equally interesting to focus on the evolution of these invaders as they respond to the CRISPR-Cas system.  Here we highlight our recent work, as well as that of others, that seeks to understand phage mechanisms of CRISPR-Cas evasion and conditions for population coexistence of phages with CRISPR-protected prokaryotes.  We begin with a look at the nature of the host cell's defense that puts pressure on phages to diversify.  Then, we describe the theoretical conditions and advantages of phage protospacer evolution, followed by experimental observations of this as well as observations of novel phage counterattack mechanisms.  We end with a couple of representative clever applications that utilize the phage-CRISPR host interaction.

\section{Targeting of phages by CRISPR}

\subsection{CRISPR spacer content}

The CRISPR spacer content provides a record of the phages to which
bacteria have been exposed, as viewed through the lens of
selection.  Experiments with \emph{S.\ thermophilus}~\cite{31} and
\emph{Leptospirillum}~\cite{56} have shown that the diversity of
CRISPR spacers in a population of bacteria decreases with distance
from the leader.  Many subsequent studies confirmed these initial
observations.  However, some studies showed a more
uniform dependence of diversity with distance from the leader.

In one of the first theoretical studies of the CRISPR system,
we sought to explain these observations using a population dynamics model~\cite{83}.  Each bacterium had a CRISPR locus of a
 finite length, with the oldest spacer dropped when the number of
spacers exceeded 30 per locus.  The CRISPR locus was 
copied to daughter cells after bacterial division.
We found that the diversity of the spacers decreased with distance
from the leader. Spacers leading to resistance
against the dominant phage were especially selected for
and accumulated in the CRISPR array.  

\begin{figure}
\begin{center}
\includegraphics[width=0.8\textwidth]{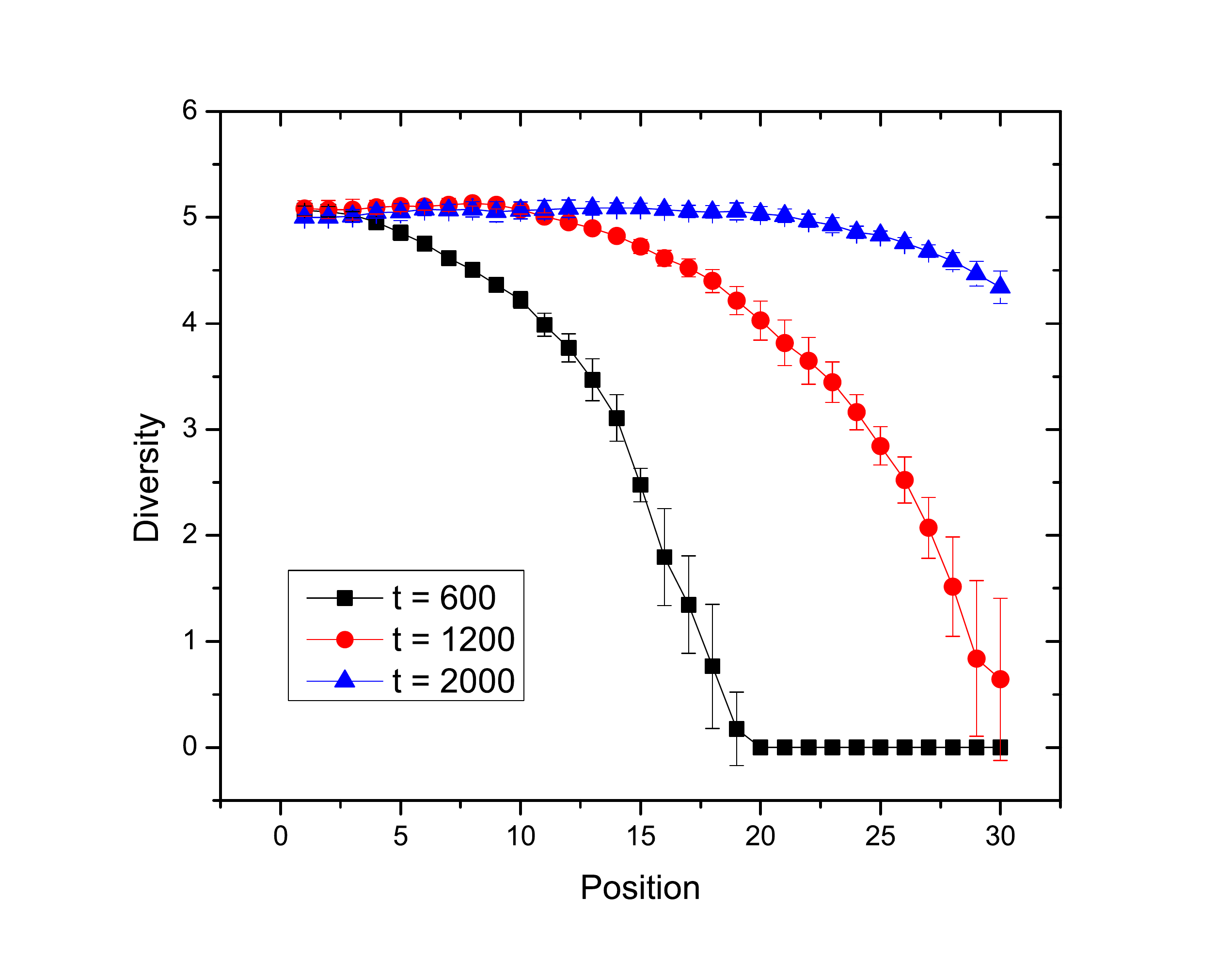}
\caption{Theoretical results for the diversity of spacers in the CRISPR locus as a function of distance from the leader sequence.  The leader-proximal spacers are more diverse than the leader-distal spacers as the CRISPR samples a new environment.  After a long time in a stable environment, the diversity of spacers becomes constant along the locus, a function of the relatively constant diversity of phages in the environment.  Reprinted with permission from~\cite{8}.}
\label{fig:spacerdiversity}
\end{center}
\end{figure}

In a second model, we sought to explain the time dependence of
this decay of diversity with distance from the leader~\cite{8}.  
Again, we found that
spacer diversity decreased towards the leader-distal end due to 
selection pressure on shorter timescales, as shown in Figure~\ref{fig:spacerdiversity}.  On longer timescales, we found
that spacer diversity was nearly constant with distance from the
leader.  Thus, spacer diversity decays more rapidly when bacteria are
exposed to new phages, either through bacterial migration or phage
influx.  These results offer one explanation for the two differing
experimental observations of spacer diversity.

\subsection{Gain or loss of immunity}

Immunity to phages that CRISPR confers upon bacteria
is not perpetual.  Changes in the phage population
lead to abrogation of the protection
afforded by the CRISPR spacers.
Defining the spacer effectiveness as the match between a spacer and the phage strains present in a population,
we found that spacer effectiveness
 decreases towards the leader-distal end as well~\cite{8}.

While the mechanism by which protospacers from the phages
are inserted as spacers into the bacterial CRISPR array adjacent
to the leader is known, the mechanism by which spacers are deleted
is less clear.  We investigated whether the results for
spacer diversity and immunity were persistent with
changes to the mechanism of spacer deletion~\cite{8}.  The results for spacer diversity and immunity were relatively insensitive to whether the oldest spacer was deleted, one of the older spacers was deleted with increasing probability toward the leader-distal end, or a random spacer was deleted from anywhere in the locus.  This insensitivity to deletion mechanism results because selection provides a strong
bias for successful deletion of the old spacers that no
longer match actively infecting phage.

Loss of immunity can lead to oscillations in the population
size of bacteria  and phage.
This phenomenon was investigated
in a minimal, Lotka-Volterra type predator-prey model of a host with a 
heritable, adaptable immune system, \emph{e.g.}, a CRISPR-Cas system~\cite{48}.
When the immunity decay rate is larger than the immunity acquisition rate, periodic oscillations of the
populations of immune hosts, sensitive hosts, and phages
 become larger and lead to quasi-chaotic behavior.  A similar behavior is also observed for the case in which the immunity acquisition rate is greater than the immunity decay rate, however the fraction of immune hosts is larger here.
There were critical values of the phage reproduction rate separating
the phases of stable equilibria, small periodic oscillations, and 
quasi-chaotic oscillations.

\begin{figure}
\begin{center}
\includegraphics[width=0.9\textwidth]{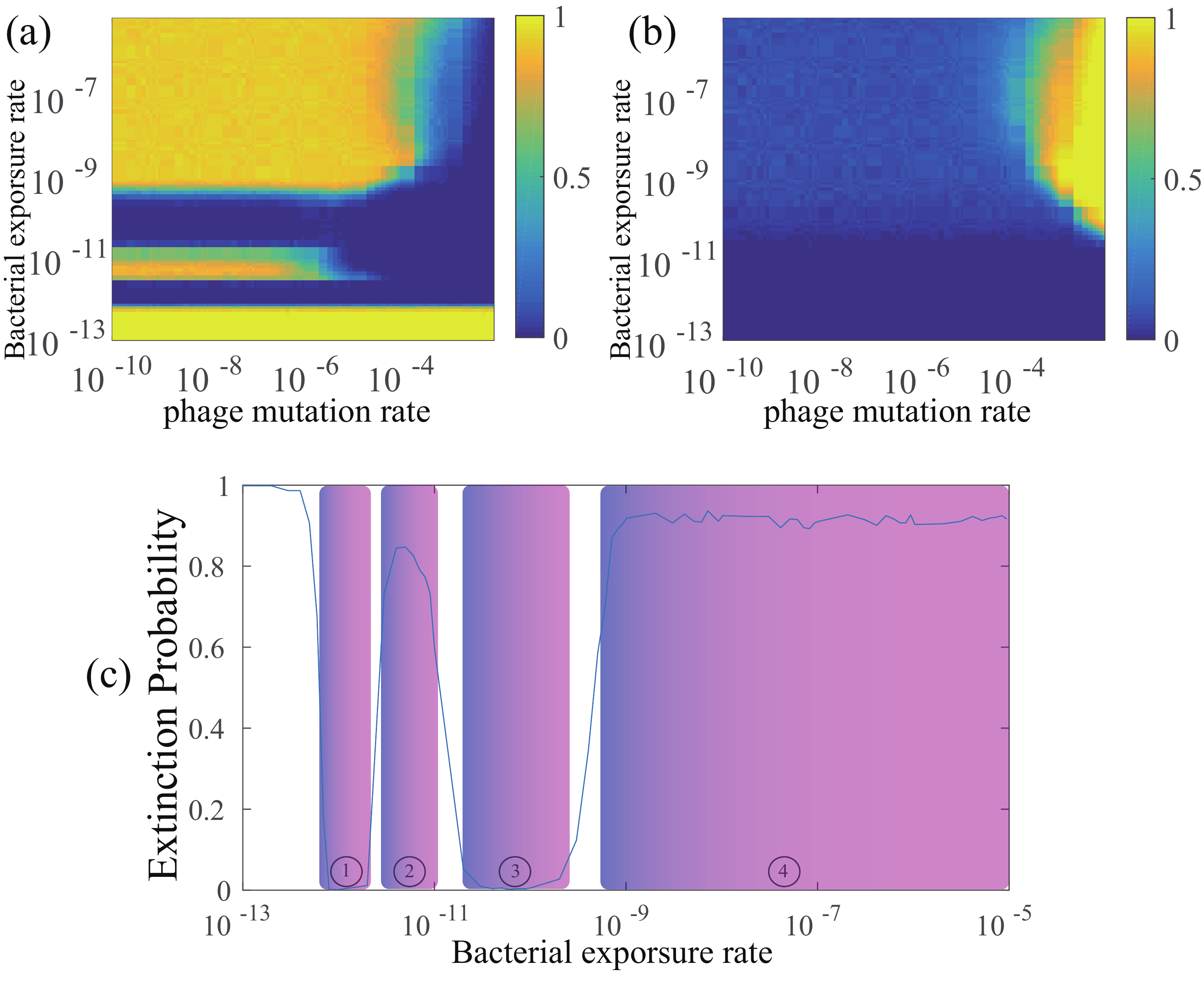}
\end{center}
\caption{
The nonclassical phase diagrams of (a) the phage extinction rate and (b) the bacterial extinction rate resulting from a coevolutionary model.  These complex patterns of phage-bacteria coexistence represent the delicate balance in place among the bacterial exposure rate, phage evasion through mutation, number of available protospacers, and rate of spacer acquisition.  (c) A small rate of CRISPR spacer deletion leads to the three observed phases of phage extinction and two phases of phage survival that depend on the rate of bacterial exposure.  Reprinted with permission from~\cite{201}.}
\label{fig:phasediagram}
\end{figure}

When the rate of spacer deletion is small, the phase diagram
no longer follows the predictions of the classical mean-field, predator-prey model.
The phage extinction probability during exposure to 
CRISPR-bearing bacteria becomes non-classical and reentrant~\cite{201}.
Parameters affecting the phase diagram include
rates of CRISPR acquisition and spacer deletion, 
rates of phage mutation and recombination, bacterial exposure rate, and 
multiple phage protospacers.
The new, non-classical region appeared at a low rate of spacer deletion, as seen in Figure~\ref{fig:phasediagram}.
The population of phages progressed through three extinction phases and 
two abundance phases, as a function of bacterial exposure rate.

\subsection{CRISPR locus length and phage diversity}

The number of spacers in the CRISPR locus and the
phage diversity are critical parameters affecting the
bacteria and phage coevolution.
In~\cite{101}, a well-defined, simple system was studied
experimentally.  Bacteria immune to a single type of phage
via a single spacer were observed to be eventually
invaded by phages.
The single spacer caused incomplete resistance because of a
high rate of CRISPR escape mutations.  That is, the bacteria
were invaded by phages that had made single
mutations in their protospacer regions.
Conversely, the CRISPR-Cas efficacy is predicted to
increase rapidly with number of 
protospacers per phage genome~\cite{54}.

Aspects of the complex phage-bacteria coevolution were also
studied theoretically.
Protection and immunity can be
non-monotonic in time
because of the decreasing phage population diversity over time
~\cite{8}.
A stochastic, agent-based mathematical model of coevolution of host and 
phage shows CRISPR-Cas efficacy is dependent on population size, spacer 
incorporation efficiency, number of protospacers per phage, phage mutation 
rate, and fitness cost of maintaining a CRISPR-Cas system~\cite{54}.
The coevolution of the CRISPR-Cas immune system and lytic phages was
modeled under evolutionary and ecological conditions, \emph{i.e.}, coupling 
of host and phage reproduction and death rates, in which 
CRISPR-Cas immunity stabilizes phage-host coexistence, rather than 
extinction of phage.  The 
overall phage diversity was observed to grow due to an increase of host and 
phage population size, not specifically due to CRISPR-Cas selection pressure on single protospacers.
The CRISPR-Cas system was predicted to become ineffective at a certain phage diversity threshold and lost due to the associated fitness cost of maintaining \emph{cas} genes.

\begin{figure}
\begin{center}
\includegraphics[width=0.8\textwidth]{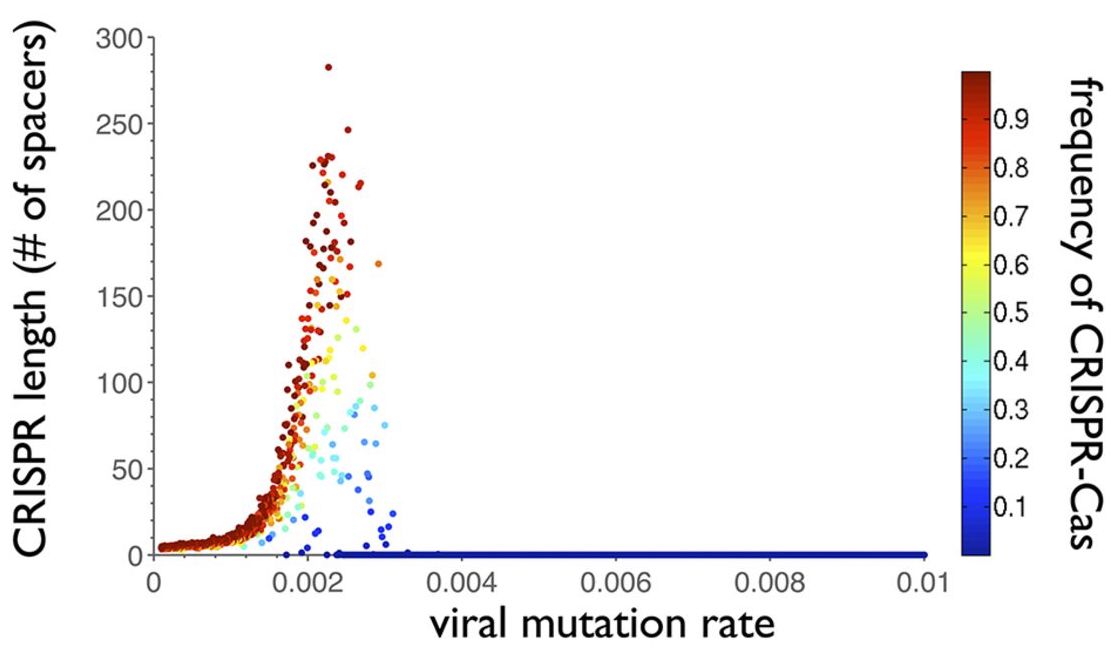}
\end{center}
\caption{
A mathematical model of how the length and prevalence of the CRISPR array depend on the phage (viral) mutation rate.  With low phage mutation rates, CRISPR-Cas systems are highly prevalent and select to retain low numbers of spacers to match the low diversity of phages.  As the phage mutation rate increases, CRISPR-Cas systems become less frequent, though those that are present are collecting more spacers to keep up with the diversifying phage population.  At a certain phage mutation threshold, the CRISPR locus becomes too long to be effectively maintained and is rapidly lost from the host population.  Reprinted with permission from~\cite{136}.}
\label{fig:lengthVdiversity}
\end{figure}

Another model similarly showed the evolved average number of spacers in the CRISPR depended on the
phage mutation rate and the spacer cost to fitness~\cite{136}.
At low mutation rates, a limited number of spacers was
sufficient to confer protection to the bacteria
against the phage population of limited diversity, shown in Figure~\ref{fig:lengthVdiversity}.
As the phage mutation rate increased, the CRISPR
loci increased in length.
At a critical threshold of phage mutation,
the CRISPR array became unable to 
recognize the diverse phage population, and the
average locus length fell rapidly to zero,
even if the rate of spacer addition outpaced phage mutation rate.
It was speculated that 
similar behavior would
occur from an increasing immigration rate of new phages.

One hypothesis for the greater fraction of hyperthermophiles that have effective CRISPR-Cas systems compared to mesophiles is that the lower rates of mutation and fixation in thermal habitats lead to more effective, and therefore selected for, CRISPR systems in thermophiles.  Additionally, another possible mechanism suggested theoretically is that CRISPR becomes ineffective in mesophiles because of larger population sizes~\cite{54}.

\section{Selection for mutation and recombination in the phage}

The bacterial immune system of CRISPR implies a selective advantage
for those phage with mutations in the PAM or protospacer regions.
That is, a mismatch between the crRNA sequence and PAM or protospacer of 
invading phage is likely to allow the phage to infect and replicate
in the bacteria.  Concomitantly, the mechanism of recombination
can integrate multiple point mutations, increasing the chance of a mismatch
that would allow the phage to escape CRISPR recognition.
In this setting, recombination can be
a positive mechanism for generating genomic diversity.

\subsection{Coevolutionary implications}

A number of coevolutionary dynamics models have captured the idea that not only can phages evade CRISPR-Cas via mutation or recombination of protospacers, but also bacteria can regain immunity through acquiring more spacers from the same phage.  These models are reviewed in~\cite{7}.  In our own work, we first considered CRISPR arrays with
between two and 30 spacers, considering the possibility of
phage mutation.  These refinements supported the
main prediction that the diversity of spacers was found to decrease 
with position from the leader proximal end~\cite{83}.

A combination of mathematical models, population dynamic experiments, and DNA sequence 
analyses have been used to understand CRISPR-containing-host and phage coevolutionary dynamics
in the \emph{S.\ thermophilus} CRISPR-Cas and virulent 
phage 2972 model systems~\cite{101}.
There was a particular interest in hosts that had gained resistance by the addition 
of novel spacers and phages that evaded resistance by mutation 
in their matching sequences.
The coevolution between the phage and bacteria was
termed an ``arms race,'' perpetuated by the competing effects
of spacer acquisition and protospacer mutation.

\begin{figure}
\begin{center}
\includegraphics[width = 0.49\textwidth]{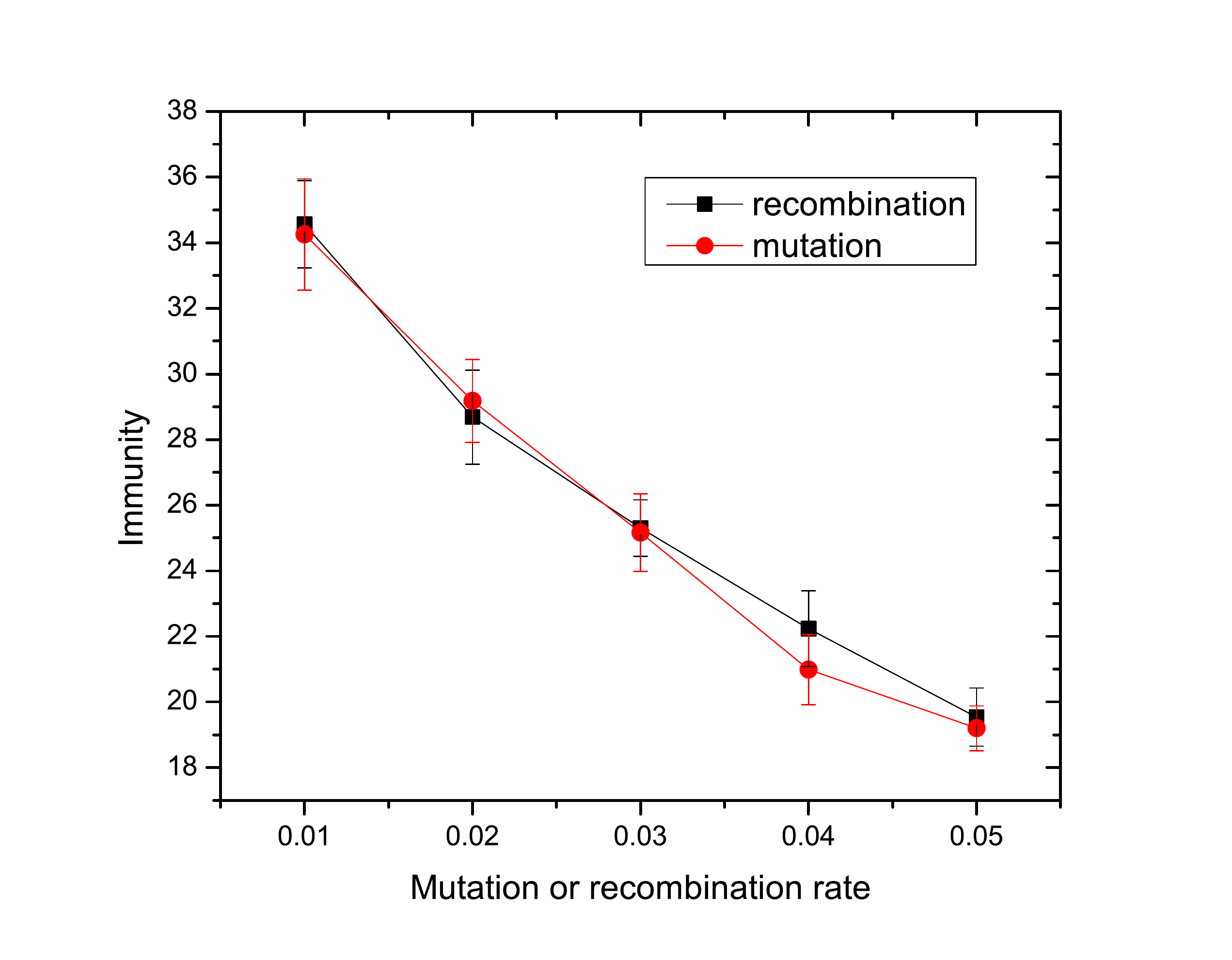}
\includegraphics[width = 0.49\textwidth]{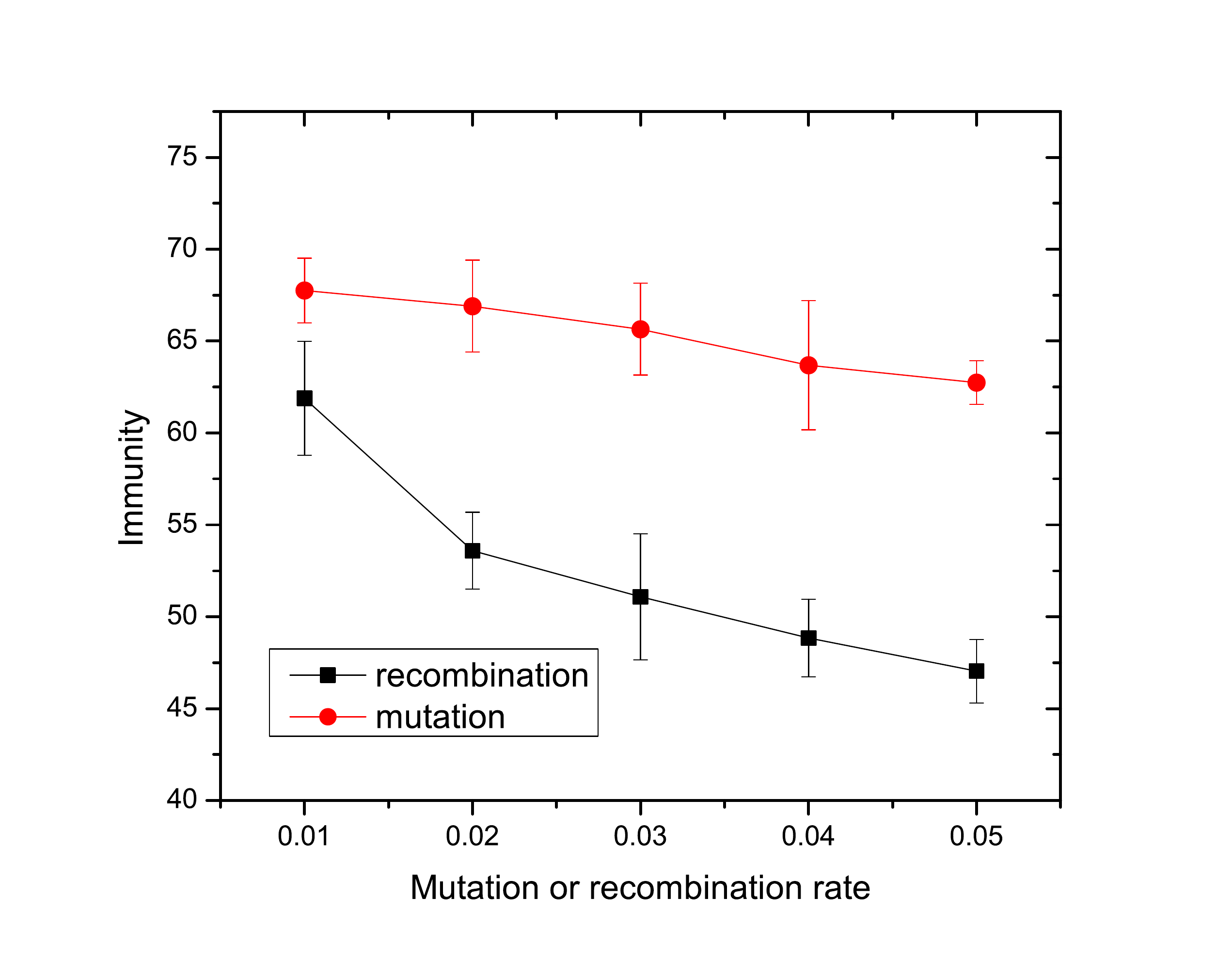}
\end{center}
\caption{
A mathematical model shows that bacterial CRISPR immunity decreases with increasing phage mutation or recombination rate.  When the bacteria's CRISPR-Cas system has a mismatch tolerance of just one nucleotide, there is little difference between the effect of phage mutation and recombination (left).  However, when there is a higher mismatch tolerance of two nucleotides, recombination gives the phage a higher probability to survive (right).  Reprinted with permission from~\cite{8}.}
\label{fig:mutationVrecombination}
\end{figure}

The effects of recombination depend on the degree of divergence
between protospacer and spacer required for the phage to escape
CRISPR surveillance.  We showed there
is little difference between the results from point mutation 
and those from recombination in the phage genome 
if the phage needs just one mismatch to escape
\cite{8}.  However, when the phage needs two mismatches to escape, the difference is apparent in the immunity, the rate at which bacteria are able to kill phages.
Recombination, by combining mutations, is a more
rapid generator of protospacer diversity and is a
more successful phage escape strategy when CRISPR has a higher 
mismatch tolerance with the protospacers, see Figure~\ref{fig:mutationVrecombination}.
When the phage has multiple protospacers, a similar argument implies
that recombination, now of the  protospacers rather than of genetic
material within a single protospacer, again leads to a more
rapid escape of the phage than does point mutation alone.
This result occurs because mutation in different
protospacers can be recombined, making it substantially
less likely for the CRISPR to recognize the recombined daughter
phage.
Thus, the phage recombination-mediated escape mechanism is also
more successful when the phage has multiple protospacers.
The immunity afforded by CRISPR is lower as mutation and recombination rates increase.

The interplay between the CRISPR pressure on the phage and the
phage pressure on the bacteria leads to a phase diagram
of coexistence.  That is, only in some parameter regions do the
phages and bacteria coexist.  The pattern of coexistence is more
complicated than the classical predator-prey model, due to the
feedback of the CRISPR system on the phage.
A low phage mutation rate can lead to a
phage extinction probability that is a non-monotonic function 
of bacterial exposure rate~\cite{201}.
The resulting nonclassical phase diagram in Figure~\ref{fig:phasediagram} shows the extinction and abundance tipping points that result from a complex relationship 
between the bacterial exposure rate and the phage mutation rates.

\subsection{Constraints on non-synonymous mutations}

One important difference between thermophiles, with habitats of $42-122^\circ$C, and mesophiles, with habitats of $20-45^\circ$C,
is that
protein stability is a more crucial factor in thermophiles.
That is, mutations are more likely to be deleterious in thermophiles,
because on average proteins are more easily destabilized
by mutation at higher temperatures
\cite{209}.
A stochastic model of phage-CRISPR coevolution was
used to investigate why CRISPR-Cas systems are more prevalent in 
thermophiles than mesophiles~\cite{136}.
Mutations are more likely to be lethal in thermophilic environments because high temperatures reduce protein stability, therefore there is selection for phages that mutate less.  It was argued that the reason
roughly 90\% of archaea, which are mostly thermophiles, have CRISPR-Cas 
systems is because they work well in this habitat.
Further support for this hypothesis was that the CRISPR-Cas is more 
correlated with thermophilic environments than with archaeal taxonomy.
This stability criterion was used to compute a
phage mutation rate threshold, beyond which
phage were selected against.

\subsection{Benefits of CRISPR vs other immune mechanisms}

Bacteria have other, innate mechanisms of resistance against
phages.  For example, the bacterial surface receptors that
promote phage attachment and entry can undergo modification.
A model was used to study the
evolution of CRISPR-Cas positive and negative hosts as they 
encountered phages~\cite{136}.
The interaction events were either successful microbial protection against infection or  successful phage infection.
The CRISPR-Cas positive host could delete or lose the CRISPR system and the
CRISPR-Cas negative host could acquire a CRISPR system by HGT. 
The fitness of the phages increased by productively infecting hosts
and creating phage progeny, whereas the fitness of the hosts increased by
acquiring CRISPR-Cas and useful spacers.  There was a potential fitness
cost to the hosts due to autoimmunity of the CRISPR system inhibiting normal
bacterial gene function and restriction of potentially beneficial
HGT events.  The CRISPR-Cas system was found to be beneficial
at intermediate levels of the innate resistance.  There was little fitness advantage from CRISPR storing 
spacers of phages that the bacteria were unlikely to encounter again.
When the bacteria survived two-thirds of its phage encounters without the help of 
CRISPR, maintaining the CRISPR system was too costly, 
\emph{i.e.}, there was no benefit to having it.  

\subsection{Heterogeneous environments}

Many of the models of the CRISPR system assume a mean-field,
homogeneous distribution of the phage and bacteria
in space.  Spatial effects and heterogenous environments, however,
occur in the body and in nature, and these effects can have
a significant effect on the outcome.
Indeed, experiments carried out with \emph{S.\ thermophilus} and phage 2972
have shown that a small percentage of acquired spacers matched the closely related phage 2766 
that had migrated spatially~\cite{58}.
A mathematical model of bacteria-phage coexistence was
used to take into account the effects of space on species coexistence and 
adaptive CRISPR defense~\cite{88}.
In the model, bacteria and phage populations spread on a two-dimensional 
square. Parameters of the model included
the effective infection rate of microcolony of bacteria,
 \emph{i.e.}, the probability of infection, and the mean latency time, 
as a ratio of phage to bacteria mean replication.
Two spatial arrangements of phage replication were explored: a well-mixed 
system, in which phage offspring were placed at random sites, and a
slow diffusion model, in which phages only spread to neighboring sites.
Bacteria dynamically acquired resistance through CRISPR-Cas to the diverse phage population while removing the oldest spacers.
For successful phage survival, \emph{i.e.}, not leading to depletion of bacterial hosts or exceeding available spatial carrying capacity, a balance was needed between the effective infection and phage replication rates.
At least two phage strains were needed to allow stable coexistence with bacteria.
Coexistence persisted as long as the maximum number of CRISPR insertions 
was fewer than the total number of phage types.

Another strain-level model of the origin and diversification of CRISPR arrays in host and protospacers in phages
was developed, taking density-dependent ecological dynamics into account
\cite{86}.  To understand the coevolution of strain diversities as well as densities, 
three main components in the model were specified: coupling among host and phage 
reproduction and death rates, molecular scale CRISPR events
  based on sequence matches between spacers and protospacers, 
and evolutionary changes of phage protospacer mutation and CRISPR spacer acquisition.
A maintenance of many coexisting strains in highly diverse communities 
was observed, with high strain similarity on short timescales but high dissimilarity over long timescales.
Short term changes in host diversity were driven by incomplete sweeps of 
newly-evolved high-fitness strains in low abundance, 
recurrence of ancestral strains that gained fitness advantage in low abundance,
and invasions of multiple dominant coalitions that arose from having nearly 
identical immune phenotypes, but different genotypes, \emph{i.e.}, similar protection afforded by the incorporation of different protospacers.
A majority of new phage mutants did not have a
significant increase in fitness, since mutation was random.
In this model only the first spacers were important to shaping selective 
coevolution because they provided the highest immunity, and the 
predicted spacer acquisition rate was more important to diversification 
than was CRISPR immunity failure.

\section{Experimental evidence for phage evasion of CRISPR}

\subsection{Synonymous mutations}

The evasion of CRISPR-Cas by phage evolution has been explored in a number of studies.  The CRISPR-Cas system creates an evolutionary battle between phage and bacteria 
through addition or deletion of spacers in the bacteria and 
mutations or deletions in the phage genomes, as reviewed in~\cite{82}.
There is evidence for phage evasion:
a small population of virulent phage mutants was observed to infect 
previously bacteriophage-insensitive bacteria. 
These infecting phages had single nucleotide changes or deletions within their targeted protospacer.
Indeed, the CRISPR locus is subject to dynamic and rapid evolutionary 
changes driven by phage exposure, and therefore CRISPR spacers can
be used to analyze past host-phage interactions.

A study of the lactic acid bacteria \emph{S.\ thermophilus} 
and the role of its CRISPR1 locus in phage-host interactions  was 
carried out by selecting two different bacteriophage-insensitive 
strains and exposing them to other phages to which they were sensitive~\cite{31}.
The addition of one new spacer of about 30 nucleotides in CRISPR1 was the most frequent outcome of a
 phage challenge.  Spacers were only acquired from protospacers with an
AGAAW motif 2 nucleotides downstream from the protospacer.
There 
was also evidence of spacer deletion. Successive phage challenges and subsequent 
addition of spacers increased the overall resistance of the host
to the phage.
Newly added spacers were required to be identical to the
protospacer region in the phage genome to confer resistance to phage.
Indeed, phages were able to evade CRISPR immunity through single nucleotide 
mutations and deletions.  The most common mechanism of escape
was mutation within the protospacers or AGAA flanking sequences.

Natural hot springs provide a rich source of microbial and phage diversity.
A metagenomic analysis of the Yellowstone hot springs to study natural 
evolution of microbial and phage populations due to the CRISPR immune system
was carried out~\cite{85}.
Two thermophilic \emph{Synechococcus} bacteria isolates 
were sequenced from microbial mats in the Octopus and Mushroom springs 
in Yellowstone National Park, \emph{Syn} OS-A from high temperature areas and \emph{Syn} OS-B' from low temperature areas, to make comparisons with 
phage and prokaryotic metagenomic data collected from the same springs.
The \emph{Syn} OS-B' genome contained individuals from two types of CRISPR 
loci, while \emph{Syn} OS-A contained individuals from three types, with
these types distinguished by their repeat sequence.
The Type III repeat sequence identified in \emph{Syn} OS-A, but not 
\emph{Syn} OS-B', was also present in another abundant microbe in the 
mat, \emph{Roseiflexus} RS-1, suggesting recent DNA transfer between them.
While CRISPR repeats were of course 
highly conserved within the microbial metagenome, 
spacer sequences were quite unique, and from the CRISPR sequence data
it was difficult to find matches to the phage metagenomic data.
Nonetheless, several spacers matched lysozyme and lysin protein genes.
Note that lysozyme enzymes attack the
cell wall late in phage infection, causing cell lysis and release of phages.
There were some silent or conservative mutations found in these phage lysozyme
or lysin protein sequences that did not affect protein function, but most likely helped the phage to evade CRISPR identification.

\subsection{Recombination}

\begin{figure}
\begin{center}
\includegraphics[width=\textwidth]{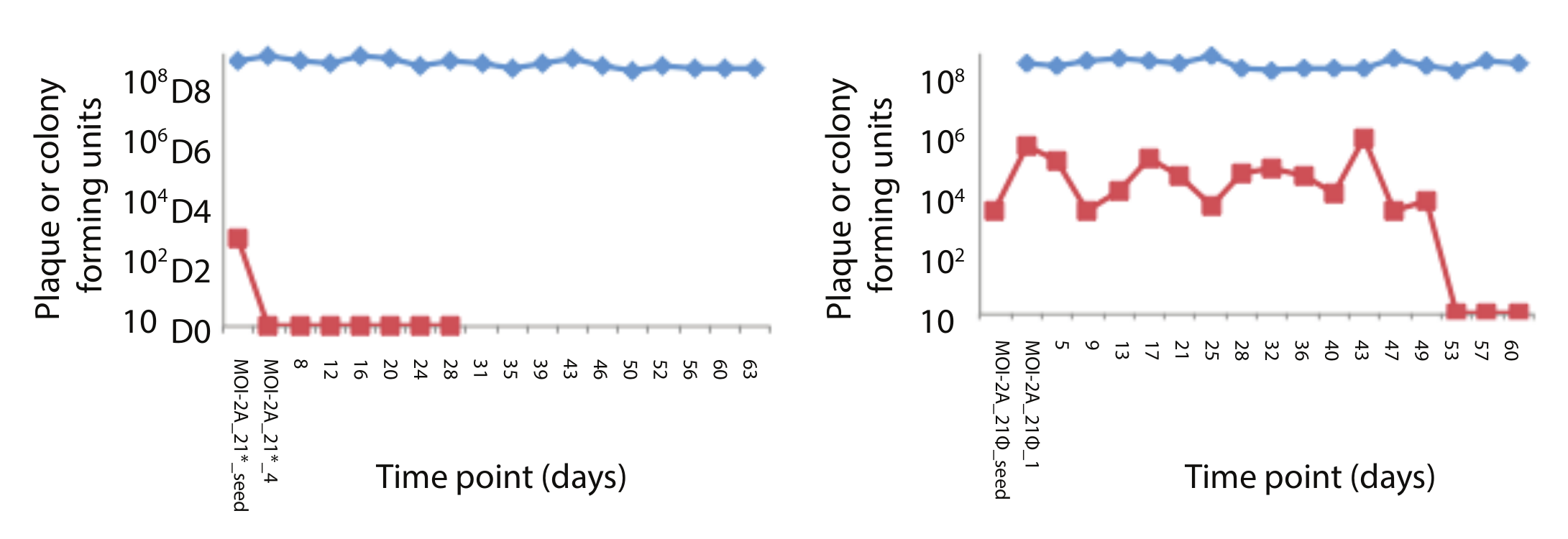}
\end{center}
\caption{
CRISPR more effectively eradicates less diverse phage populations.  Experimental evidence of the ability of two phage types to coexist with CRISPR bacteria (right) for a longer period of time than a single phage type could (left).  The host cell population (CFU per milliliter) is shown in blue, and phage counts (PFU per milliliter) are shown in red.  Reprinted with permission from~\cite{58}.}
\label{fig:extinctionVdiversity}
\end{figure}

The phage genomic regions targeted by CRISPR are driven to have mutations
\cite{58}.
Homologous recombination events between genetically related phages
can then further diversify the phage population.
Long-term coevolution experimental studies
were carried out with \emph{S.\ thermophilus} and phage 2972 
for up to 232 days, until the phage went extinct.
An analysis of the \emph{S.\ thermophilus} spacers revealed that spacers 
acquired during the experiments mapped unevenly over the phage genome.
The immune pressure from CRISPR drove escape mutations located exclusively in the protospacer and PAM regions, as well as the accumulation of phage genome rearrangements.  Phage mutation rates were much higher than that of the
bacterial host, and the presence of phages also accelerated host genome 
evolution in the CRISPR array.
The coexistence of multiple phages allowed recombination events that
boosted the observed substitution rates beyond the bare
mutation rate, termed the ``rescue'' effect, see Figure~\ref{fig:extinctionVdiversity}.

Experiments show that the most recently acquired spacers match coexisting 
phages~\cite{62}.  In other words, the phage population evolves,
abrogating the utility of old spacers.
Phages use extensive recombination to shuffle sequence motifs and 
evade the CRISPR spacers.
Spacers of CRISPR loci recovered from \emph{Leptospirillum} group III, 
I-plasma, E-plasma,  A-plasma, and G-plasma microbial communities in biofilms collected 
from Richmond Mine (Redding, CA) were used to link phages
to their coexisting host 
bacteria and archaea.
A majority of spacers corresponded to phages, though some corresponded to 
other mobile genetic elements, such as plasmids and transposons.
It was found that microbial cells targeted several different phage populations,
 and only a few CRISPR spacers were widely shared among bacterial strains. 
For example, some E-plasma cells target specific AMDV2, AMDV3, and AMDV4 phage 
variants, and some spacers match dominant phage sequence types. 
Yet many spacers match sequences characteristic of only one or a small number of genotypes within the population.
The phage population is reshaped by extensive homologous recombination, 
as evident from combinatorial mixtures of small sequence motifs. 
This recombination resulted in genetic blocks shared by different phage
individuals that were often no more than 25 nucleotides in length, creating 
new phage sequences that can disrupt the function of the CRISPR's 28 to 
54-nucleotide spacers.
A benefit of recombination is that if the preexisting sequence
diversity in the phage population is mainly in the
protospacer and PAM regions, recombination can increase
this diversity in a combinatorial way without introducing novel diversity
outside these regions.
In this fashion,
recombination creates new DNA sequences with 
a lower risk of altering protein function than does mutation and limits alterations to the phage genome outside of the CRISPR-recognized PAMs and protospacers.

\subsection{Mutations in PAM, seed, and non-seed regions are distinct}

The original view that a single mutation in the protospacer region is
sufficient to allow phage to escape from CRISPR has transformed into a more
nuanced view of CRISPR recognition being quite sensitive to mutations in
a seed region, which is usually defined as the protospacer's eight PAM-proximal nucleotides~\cite{55}, but less sensitive to mutations in the rest of the protospacer region.
An experiment with a co-culture of 
\emph{S.\ thermophilus} with phage 2972 for one week 
studied the impact of spacer acquisition and host population diversification
on phage genome evolution ~\cite{57}.
Tracking of CRISPR diversification and host-phage coevolution
revealed a strong selective advantage for phages containing PAM or near PAM mutations.
A genetically diverse bacterial population arose, 
with multiple subdominant strain lineages. 
All surviving \emph{S.\ thermophilus} cells had at least one newly 
incorporated spacer against phage 2972.
Of the two loci sequenced, CRISPR1 was the most active, and
CRISPR3 only incorporated single spacers.
In this experiment, 
all recovered phages contained a synonymous mutation eight nucleotides 
from the PAM, apparently leading to escape from
\emph{S.\ thermophilus} spacer1 in the CRISPR1 loci.  The fixation of this 
mutation suggests that  only phage that had it could rise in abundance.
This phage sequence encodes a protein that recognizes the host, and
there is strong selective pressure for the host to 
abrogate this recognition.
Also in this experiment,  88\% of recovered phage
sequences contained synonymous mutations six nucleotides from PAM, 
likely leading to escape from spacer32.
Finally, 92\% of phage sequences
contained non-synonymous mutations in the PAM corresponding to spacer6.
There were no observed
phage mutations in the regions targeted by any CRISPR3 spacers.

\subsection{Long-term study of heterogeneous environments}

A long term metagenomic study of archaeal, bacteria, and phage populations 
in Lake Tyrrell (LT) from 2007 to 2010 was carried out, and the dynamics 
of their populations on timesscales of months to years was studied~\cite{65}.
Samples were collected and analyzed from LT over three summers and four winters.
Overall, archaeal and bacterial populations were
more stable than the phage populations, \emph{i.e.}, over 
the timescale of years, phage populations were less stable than their prokaryotic prey.  Analysis of CRISPR arrays indicated both rare and abundant phages were 
targeted, suggesting archaeal hosts attempt to balance protecting themselves 
against persistent, low-abundance phages and highly abundant phages 
that could destroy the host community.  There was a high diversity of phages in the environment, and even in the absence of superinfection, the CRISPR array sampled extensively from this diversity.

\subsection{Plasmid evasion of CRISPR-Cas}

While plasmids and phage bear a resemblance, interestingly the selection 
pressure on plasmids from CRISPR appears somewhat different from that
on the phage.  Namely, the rate of plasmid mutation is much slower than combined rate of loss of CRISPR immunity by spontaneous mutation or deletion.
On the one hand, plasmids can confer positive features upon bacteria,
such as antibiotic resistance.  Analysis via experiments and computer modeling of 
the loss of CRISPR-Cas loci in the presence of an environment containing 
plasmids that increase the host's fitness has been carried out~\cite{102}.
Conjugational transfer of the Staphylococcal plasmid pG0400 (\emph{nickase} gene
\emph{nes}) into \emph{Staphylococcus epidermidis}, which contained 
a spacer targeting this plasmid, was analyzed.
Simulation results showed plasmid transfer into the host could occur if
the plasmid mutated, the CRISPR lost the associated spacer, 
the CRISPR locus became deactivated or deleted, 
or the CRISPR response was subdued.
Experiments showed the wild-type plasmid on CRIPSR-negative mutants only,
 meaning that instead of utilizing the phage mechanism of mutating their targeted regions to evade CRISPR-Cas, a plasmid ``evasion'' strategy could occur within the host, \emph{i.e.}, loss of CRISPR locus allowed the host to receive the beneficial plasmid.  \emph{In vitro} experiments showed little to no intrinsic fitness cost of losing CRISPR.

\section{Emergence of game theoretic strategies in the phage}

\subsection{Anti-CRISPR proteins}

\begin{figure}
\begin{center}
\includegraphics[width=\textwidth]{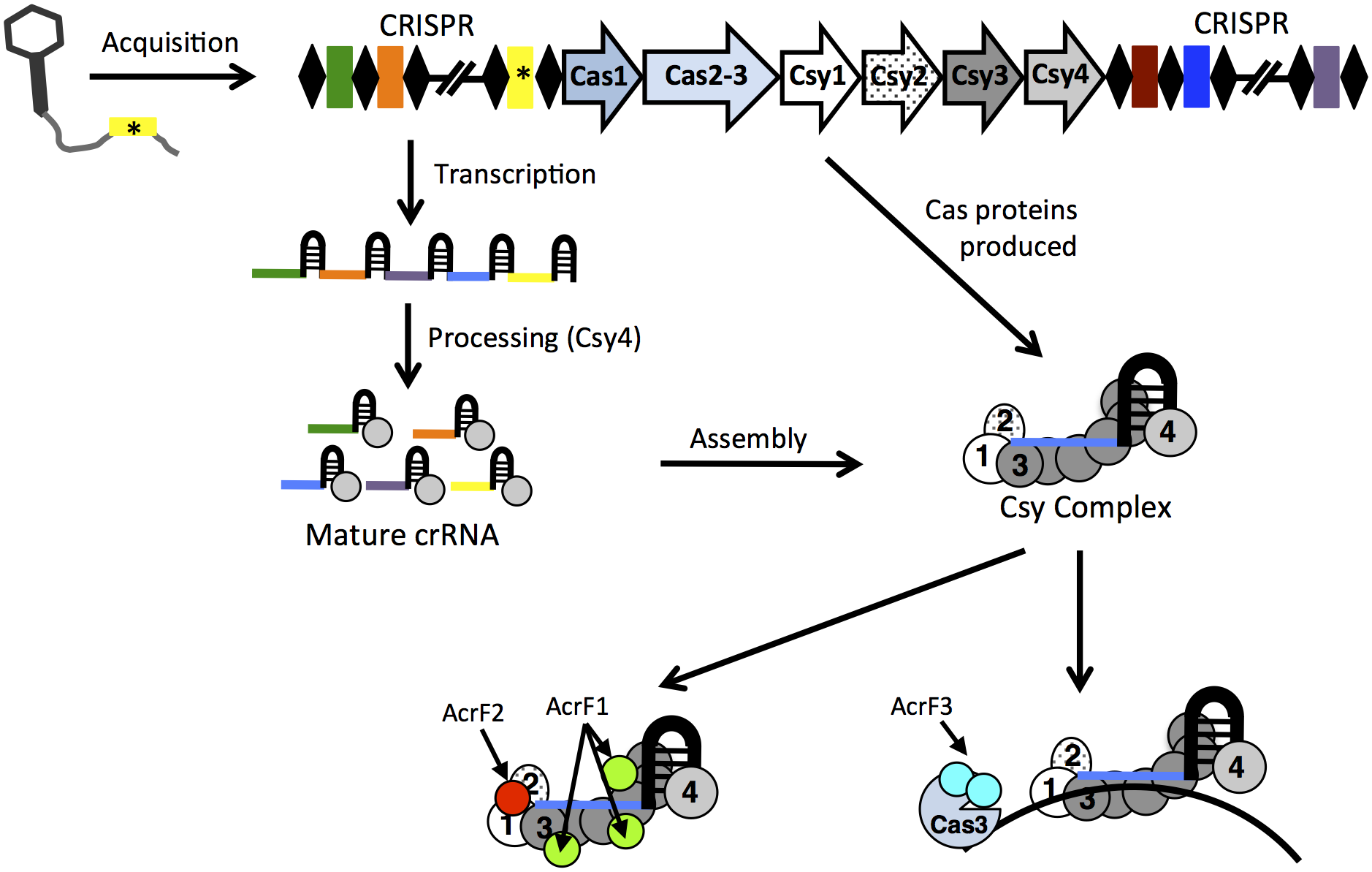}
\end{center}
\caption{
Depiction of mechanisms used by three different anti-CRISPR proteins that target the Type I-F CRISPR system, which contains a crRNA:Csy surveillance complex and Cas3 cleavage protein.  Anti-CRISPR proteins AcrF1 and AcrF2 attach directly to the the Csy complex to block the crRNA from fully binding with the target DNA.  Protein AcrF3 attaches to the Cas3 protein to prevent it from being recruited to cleave a bound target. Reprinted with permission from~\cite{192}.}
\label{fig:antiCRISPR}
\end{figure}

The phages evade CRISPR by more than just mutation and recombination.
Phages have evolved more complicated mechanisms that may be
understood from a game theoretic point of view.
Anti-CRISPR proteins have been identified
in phage that are associated with inactivating Type I-E and I-F CRISPR-Cas 
systems~\cite{192}.
These anti-CRISPR proteins, discovered to be
 encoded by \emph{Pseudomonas aeruginosa} phages, circumvent CRISPR by inactivating the Cas proteins.  Five distinct families of proteins that 
targeted Type I-F and four that targeted Type I-E were found.
The existence of subsets of these genes among phages suggests
HGT may have been responsible for a
``mix and match'' scenario of acquiring them.
The mechanisms of action of three unique I-F interference inhibitors, AcrF1, AcrF2, and AcrF3,
are illustrated in Figure~\ref{fig:antiCRISPR}.
Briefly, in Type I-F CRISPR-Cas systems, a Csy complex is guided by crRNA to bind to invading DNA, and Cas3 is recruited for target cleavage.
 AcrF1 binds along the full Csy3, which comprises three molecules in the Csy complex, but allosterically interferes with DNA binding. AcrF2 binds to the Csy1-Csy2 heterodimer in the Csy complex to block the 5' end of crRNA and directly prevents DNA binding.
AcrF3 interacts with Cas3 to block its recruitment to the Csy complex.
These mechanisms could potentially regulate lateral gene transfer to
allow foreign DNA to bypass CRISPR-Cas inhibition.
In a separate experiment, it was also found that another
phage produced enzymes upon infection that induced a phenotypic phage 
resistance in sensitive bacteria, but killed bacteria with CRISPR, even in the presence of many phages~\cite{101}.

\subsection{Phage encoding its own CRISPR}

Amazingly, CRISPR elements have even been 
found in prophages and phage DNA segments.
Whole-genome microarray analysis revealed that
the \emph{Clostridium difficile} genome contained mobile genetic elements, 
such as prophages, which contained CRISPR loci~\cite{168}.
The complete genome sequence was
 determined for strain 630 of \emph{C.\ difficile}, which is virulent 
and multidrug-resistant.
An unusually large fraction of the genome,
11\%, consists of mobile genetic elements (MGEs), including two prophages 
and a prophage-like element, responsible for acquisition of genes involved 
in antimicrobial resistance and virulence.
Ten CRISPR DNA repeat regions were
identified with no evidence of their expression or function. Interestingly, 
several were located on the prophages and prophage-like element, suggesting the
phages had incorporated a CRISPR with spacers against other phages.

Twenty-two CRISPR arrays were found in the phage sequences
from a metagenomic study of the genetic composition of the phage population in the human gut microbiome~\cite{200}.  Simultaneously isolated at two time points, one of the phage's spacer sequences matched the sequence of another phage present in the same individual, suggesting a phage-phage competition mediated by CRISPR.  Additionally, there was an array that showed greater than 95\% identity in the repeat regions to the previously found CRISPR in \emph{C.\ difficile}~\cite{168}, described above.

One bacteriophage has been observed to directly combat a host bacteria's immunity by using a CRISPR-Cas system~\cite{161}.
The bacterial \emph{Vibro cholerae} serogroup O1, which causes cholera, 
can be treated with the {\bf I}nternational {\bf C}entre for Diarrhoeal Disease Research, Bangladesh cholera phage 1 (ICphage1).
The currently active \emph{V.\ cholerae} strain ``El Tor'' does not contain a CRISPR-Cas system, however it encodes an 18kb {\bf p}hage-inducible chromosomal island-{\bf l}ike {\bf e}lement (PLE).  The phage-inducible chromosomal island is a highly mobile genetic element that contributes to HGT, host adaptation, and virulence, by using phages as helpers to promote the host's spread while simultaneously preventing these helper phages from reproducing.  \emph{V.\ cholerae} therefore uses its PLE to interfere with the ICphage1 reproductive cycle and increase its own virulence.  As an evolutionary counterattack, the ICphage1 contains a CRISPR-Cas system, comprised of two loci and six \emph{cas} genes, that actively targets the bacteria's PLE.  A single spacer that targets the PLE is sufficient to allow the ICphage1 to destroy the PLE and then replicate successfully.  It is unclear how the ICphage1 evolved to have this system, however a comparison with existing characterized CRISP
 R systems reveals a high similarity to Type I-F.

\section{Discussion}

\subsection{An example application in biotechnology}

The CRISPR system has powerful applications in the
molecular biology and genomic editing fields.  In the latter, CRISPR has been used to
delete, add, activate or suppress targeted genes in many species, including
human cells, resulting in the so-called ``CRISPR craze''~\cite{D}.
Self-targeting applications, to target and cleave the host chromosome, have
been suggested.  These include antimicrobial selection for specific microbial populations within a mixture,
 antibiotic resistant gene targeting, and large scale genomic deletion~\cite{4,83}.
Targeting of the uptake of external genetic material to inhibit the spread of
resistance genes by HGT has also been suggested~\cite{83}.
Here we mention one successful experimental implementation of using phage-delivered CRISPR-Cas to control antibiotic resistance and horizontal gene transfer~\cite{162}.  In this study, temperate lambda phages were used to deliver a CRISPR-Cas system into \emph{E.\ coli}.  The locus was programmed to destroy antibiotic resistance-conferring plasmids and protect against lytic phages.  Following its delivery, lytic phages were used to kill antibiotic-resistant bacteria. The combined result was to leave only antibiotic-sensitive bacteria, which could later be killed via antibiotics. This method selected for antibiotic-sensitized bacteria by creating and delivering to the bacteria a CRISPR-Cas system that contained spacers which would first self-target the antibiotic-resistance genes and then protect the bacteria from lytic phages engineered with matching protospacers.  It was found that HGT of antibiotic resistant elements was  no longer possible to \emph{E.\ coli} containing the
  system. Additionally, these antibiotic-sensitive bacteria were resistant to the engineered lytic phages.  Importantly, the construction
allowed for the selection of these bacteria and selection against antibiotic-resistant ones.  This bacteria-sensitizing method delivered a CRISPR-Cas system to phages that then delivered this system to bacteria, which simulated relevant external environments such as a hospital fomites or the human skin.  An important advantage, therefore, is that this approach would not require delivery of the CRISPR-Cas system to human host cells.

\subsection{An example application in microbiome modification}

Study of the microbiome is a burgeoning field, with implications ranging from
health and disease to learning and mood.  The bacterial flora in the gut
are heavily influenced by the 15$\times$ greater number of viruses there~\cite{210}, 90\% of which are bacteriophages~\cite{211}.
A study of the genetic composition of the phage population in the human gut 
and their dynamic evolution in response to environmental perturbations
was carried out~\cite{200}.
Metagenomic sequencing of the human virome from subjects on a dietary 
intervention of a high-fat and low fiber, a low-fat and high fiber, or an ad-lib diet
was performed on samples collected over eight days
 to understand the structure and dynamics of the phage population.
The controlled feeding regimen caused the virome to change and converge 
among individual subjects on the same diet.
The sequencing identified many DNA segments responsible
for phage functions required in lytic and lysogenic growth, 
as well as antibiotic resistance.  Interestingly, the study found phages encoded CRISPR against other phages, as reported in the previous section.
CRISPR arrays in \emph{Bacteriodetes} from the gut contained
spacers matching genetic material from the virome sequenced 
in the same individuals, suggesting a functional CRISPR 
existed.

\section{Conclusion}

CRISPR-Cas is now a major topic in applied molecular biology and genomic editing.  It first rose to prominence, however, as a novel immune defense mechanism of bacteria against phages.  Theory and modeling have shed light on the dynamics of CRISPR spacer acquisition, how the diversity of the phage population is reflected in the content of the spacers, and the potentially complicated patterns of coevolution that bacteria and phages exhibit.  CRISPR puts selection pressure on phages to evolve, selecting for increased rates of substitution and recombination.  The effect that heterogeneous environments have on phage-bacteria coexistence has also been explored.  A number of experimental studies have borne out theoretical predictions regarding phage diversity, mutation and recombination, and heterogeneous environments.  Interestingly, phages also encode strategies to combat the bacterial immune system, including anti-CRISPR proteins and their own CRISPR-Cas system.  Theoretical wor
 k and modeling continue to play an integral role in developing a fundamental understanding of the CRISPR-Cas genetic adaptive immune system of prokaryotes and in designing applications in molecular biotechnology and genomic editing.

\section*{Acknowledgement}
This work was partially supported by the Center for Theoretical Biological Physics at Rice University, Houston, TX 77005, USA.

\bibliographystyle{unsrt}
\bibliography{DeemCRISPR}{}

\end{document}